\documentclass[12pt,epsf]{article}

\usepackage{epsfig}
\usepackage{amssymb}
\usepackage{graphicx}
\usepackage{color}

\begin{document}

\baselineskip 6mm


\newcommand{\nc}{\newcommand}
\newcommand{\rnc}{\renewcommand}

\headheight=0truein
\headsep=0truein
\topmargin=0truein
\oddsidemargin=0truein
\evensidemargin=0truein
\textheight=9.5truein
\textwidth=6.5truein

\rnc{\baselinestretch}{1.24}    
\setlength{\jot}{6pt}       
\rnc{\arraystretch}{1.24}   



\newcommand{\tcb}{\textcolor{blue}}
\newcommand{\tcr}{\textcolor{red}}
\newcommand{\tcg}{\textcolor{green}}


\def\be{\begin{equation}}
\def\ee{\end{equation}}
\def\ba{\begin{array}}
\def\ea{\end{array}}
\def\bea{\begin{eqnarray}}
\def\eea{\end{eqnarray}}
\def\nn{\nonumber\\}


\def\ct{\cite}
\def\la{\label}
\def\eq#1{Eq. (\ref{#1})}


\def\a{\alpha}
\def\b{\beta}
\def\g{\gamma}
\def\G{\Gamma}
\def\d{\delta}
\def\D{\Delta}
\def\ep{\epsilon}
\def\e{\eta}
\def\ph{\phi}
\def\Ph{\Phi}
\def\ps{\psi}
\def\Ps{\Psi}
\def\k{\kappa}
\def\l{\lambda}
\def\L{\Lambda}
\def\m{\mu}
\def\n{\nu}
\def\th{\theta}
\def\Th{\Theta}
\def\r{\rho}
\def\s{\sigma}
\def\S{\Sigma}
\def\ta{\tau}
\def\o{\omega}
\def\O{\Omega}
\def\pr{\prime}


\def\half{\frac{1}{2}}

\def\goto{\rightarrow}

\def\na{\nabla}
\def\grad{\nabla}
\def\curl{\nabla\times}
\def\div{\nabla\cdot}
\def\pa{\partial}
\def\ex{{\rm e}}

\def\lb{\left[}
\def\lc{\left\{}
\def\ls{\left(}
\def\ln{\left.}
\def\rn{\right.}
\def\rb{\right]}
\def\rc{\right\}}
\def\rs{\right)}

\def\vac#1{\mid #1 \rangle}


\def\td#1{\tilde{#1}}
\def\check{ \maltese {\bf Check!}}


\def\Tr{{\rm Tr}\,}
\def\det{{\rm det}}


\def\bc#1{\nnindent {\bf $\bullet$ #1} \\ }
\def\ch {$<Check!>$ }
\def\ss {\vspace{1.5cm}}

\begin{titlepage}

\hfill\parbox{5cm} { }

\vspace{25mm}

\begin{center}
{\Large \bf On the giant magnon and spike solutions for strings on
AdS$_3\times$ S$^3$}
\vskip 1. cm
  {\large Bum-Hoon Lee$^a$\footnote{e-mail : bhl@sogang.ac.kr},
  Rashmi R. Nayak$^a$\footnote{e-mail : rashmi@sogang.ac.kr},
  Kamal L. Panigrahi$^b$\footnote{e-mail : panigrahi@iitg.ernet.in} \\
and  Chanyong Park$^a$\footnote{e-mail : cyong21@sogang.ac.kr}}

\vskip 0.5cm

{\it $^a\,$Center for Quantum Space-Time (CQUeST), Sogang University, \\
   Seoul 121-742, Korea}\\
{ \it $^b\,$ Department of Physics, Indian Institute of Technology Guwahati, \\
Guwahati-781 039, India}\\

\end{center}

\thispagestyle{empty}

\vskip1cm


\centerline{\bf ABSTRACT} \vskip 3mm

\vspace{1cm} We study solutions for the rotating strings on the
sphere with a background NS-NS field and on the Anti-de-Sitter
spacetime. We show the existence of magnon and single spike
solutions on R$\times$S$^2$ in the presence of constant magnetic field as
two limiting cases. We also study the solution for strings on
AdS$_3\times$ S$^3$ with Melvin deformation. The dispersion
relations among various conserved charges are shown to receive
finite corrections due to the deformation parameter. We further
study the rotating string on AdS$_3 \times$ S$^3$ geometry with
two conserved angular momenta on S$^3$ and one spin along the
AdS$_3$. We show that there exists two kinds of solutions: a
circular string solution and a helical string. We find out the
dispersion relation among various charges and give physical
interpretation of these solutions.

\vspace{2cm}


\end{titlepage}

\renewcommand{\thefootnote}{\arabic{footnote}}
\setcounter{footnote}{0}

\section{Introduction}
A remarkable development in the field of string theory is the
celebrated string theory-gauge theory duality, which relates the
spectrum of free string on $AdS_5 \times S^5$ with that of
operator dimension of ${\cal N}=4$ super Yang-Mills (SYM) in
planar limit. Determining this spectrum is an interesting and
challenging problem. Recently it has been realized that this
problem of counting the operators in gauge theory has an elegant
formulation in terms of {\it integrable} spin chain
\ct{zm,intg1,intg2,intg3,intg4,intg5,intg6}. In the dual
formulation, the string theory has also integrable structure in
the semiclassical limit. Recently Hofman and Maldacena (HM)
considered a special limit where the problem of determining the
spectrum on both sides simplifies considerably \ct{hm0604}. In
this limit the 't Hooft coupling $ \lambda $ is held fixed
allowing for a direct interpolation between the gauge theory
$(\lambda << 1)$ and string theory $(\lambda >> 1)$ and the energy
$E$ (or conformal dimension $\Delta = E$) and a $U(1)$ R-charge
$J$ both become infinite with the difference $E - J$ held fixed.
The spectrum consists of elementary excitations known as magnons
that propagates with a conserved momentum $p$ along the long spin
chain \ct{hm0604}\footnote{for more work on related topic see for
example
\cite{Dorey:2006dq,gmag,Bobev:2006fg,Chen:2006gq,Hirano:2006ti,Maldacena:2006rv,
Kluson:2007qu,Kluson:2007fr,Hofman:2007xp,Dorey:2007an}}. These
magnon excitations satisfy a dispersion relation of the type (in
the large 't Hooft limit $(\lambda)$)
\begin{equation} E-J = \frac{\sqrt{\lambda}}{\pi}
\left|{\rm sin}\frac{p}{2}\right| \ .
\end{equation}
A more general type of solution are the ones rotating in AdS$_5$,
one of which is the spiky string \cite{k0410} which describes the
higher twist operators from field theory view point. Giant magnon
solutions can be seen as a special limit of such spiky strings
with shorter wavelength. Several papers
\ct{ik0705,krec,jin,sv1,sv2,krt0607,oka1,oka2,tsey1,tsey2,
tsey3,tsey4,tsey5,tsey6,tsey7} have been devoted in studying the
gauge theory and string theory side of this interesting rotating
solutions in AdS space and on the sphere. Hence it is very
important to find out more general class of rotating and pulsating
strings in AdS$_3 \times$ S$^3$ background and look for possible
dual operators in the gauge theory. Because the complete
understanding of the gauge theory operators corresponding to the
semi classical string states is still lacking, it seems reasonable
to find out the string states first and then look for possible
operators in the dual side.

In this paper we study few examples of spike solutions for strings
in AdS$_3 \times$ S$^3$ background in an attempt to study the
string spectrum and the other elementary string states further.
The AdS$_3 \times$ S$^3$ spacetime has been studied by using the
$SL(2,R) \times SU(2)$ Wess-Zumino-Witten model. In the study of
D-branes on this group manifold, one needs the mechanism of `flux
stabilisation' which ensures the stability of these branes against
the collapse of the sphere. This has opened up a new window for
the understanding of strings and branes in AdS space in the past.
Our interest is whether one could find out any rotating string
solution that looks like a spike and/or magnon in this background.
As we will show in what follows that there exist such a solution,
which modifies the relation between various conserved charges of
the spike solution in a very natural way. Our next example is the
spike solution in a Melvin deformed AdS$_3 \times$ S$^3$
background, and we will show that in the limit of small {\it
deformation} parameter, to the leading order, the {\it energy Vs
height} of spike, relationship gets corrections even at the lowest
order in $\lambda$. Finally we present an interesting example of
elementary string solution with one spin along AdS$_3$ and two
angular momenta along S$^3$. We find a parameter space of
configurations which admit two interesting classes of solution.
One of them is a classical circular string on AdS$_3$ with the
infinite spin $S$ and at the same time, the giant magnon on S$^3$
with the finite angular momentum. The other is a helical string
which has the same configuration with the circular string on the
sphere but becomes an array of the spikes on AdS$_3$. We will show
that these two solutions satisfy similar dispersion relation with
two parameters, the velocity of the string $v$ and the winding
number $\td{w}$. In the absence of an exact expression for energy,
we will write a perturbative expansion form of the dispersion
relation and give the physical meaning of this solution. For the
helical string case, we will find the dispersion relation for a
single spike which is one segment of the helical string.

The rest of the paper is organized as follows. In section 2, we
calculate the energy and momentum of a spike on a two sphere with
a constant background NS-NS B$_{\mu\nu}$ field. We have shown that
for the rigidly rotating string of the two sphere, in the
background of constant NS-NS B-field, there exists two limiting
cases of interest. First one is the known magnon solution studied
in \cite{h0607,h0512}. The second one is the single spike solution that
generalizes of the single spike solution on R$\times$S$^2$ found in
\cite{ik0705}. We compute its energy E and angular momentum
$J$ and a function of $\bar\theta$ (the height of the spike) and
the constant background field. Section 3 is devoted to the study
of spike like solution in the magnetic Melvin deformed AdS$_3
\times$ S$^3$, where we constrain the motion of the string along
R$\times$S$^3$ only. In the small deformation parameter, we show that the
relationship between the angular momenta and the height of the
spike, which is a generalization of the result obtained in
\cite{ik0705}. In section 4 we calculate the example of multi
spin spike solution in the AdS$_3 \times$ S$^3$ background with
two angular momenta along S$^3$ and a spin along AdS$_3$. We find
two classes of solution of particular interest and the dispersion
relation for each. These multi-spin solutions can be reinterpreted
as a generalization of the giant magnon on S$^2$ with other spins
and have different shapes on the AdS space. Finally in section 5,
we conclude with some remarks.

\section{Spike on R$\times$S$^2$ with a background NS-NS $B$ field}

As a first example we will show the existence of a single spike
solution of the string around R$\times$S$^2$ with a background NS-NS $B$
field. We will show that for the string rotating around the rigid
sphere in the background of B field, there exists two interesting
solution. The first one is a magnon solution found in
\cite{h0607}, and the second one is a single spike solution which
generalizes the results of \cite{ik0705} can be obtained from two
different limits of the same solution. As explained in
\cite{Bachas:2000ik}, the NS-NS background field has been used for
the purpose of stability of the size of the sphere against its
shrinking to zero size. The metric and the background NS-NS flux
field is given by \bea ds^2 = -dt^2 +{d\theta}^2+ \sin^2\theta
{d\phi}^2, \>\>\>\> B_{\theta \phi} = B \sin\theta
\label{s2-b}\eea We are interested in finding out the classical
rotating string solution around this geometry. To do so, as usual
the starting point is to write down the Polyakov form of the
action \bea\label{actPol} S = -\frac{\sqrt{\lambda}}{4\pi}
\int_{-\pi}^\pi d\sigma d\tau [\sqrt{-\gamma}\gamma^{\alpha\beta}
g_{MN}\partial_\alpha x^M\partial_\beta x^N -e^{\alpha\beta}
\partial_\alpha x^M\partial_\beta x^N B_{MN}] \eea and
where $\gamma^{\alpha\beta}$ is world-sheet metric and
$e^{\alpha\beta}$ is the anti symmetric tensor defined as
$e^{01}=-e^{10} = 1$. Finally, the modes $x^M, M=1,\dots,9$
parameterize the embedding of the string in the background. The
equations of motion derived by the above action has to be
supplemented by the following Virasoro constraints \bea
&&g_{MN}(\partial_\sigma x^M\partial_\sigma x^N +
\partial_\tau x^M\partial_\tau x^N)= 0 \cr & \cr
&& g_{MN}\partial_\sigma x^M\partial_\tau x^N = 0 \eea We consider
a spike string in the following worldsheet parametrization \bea t=
\k\tau,~~~ \theta=\theta(y),~~~ \phi= \omega\tau + \tilde
\phi(y).\eea where $y = \alpha \sigma + \beta \tau$. With this the
Virasoro constraints take the form \bea &&\dot{\theta}
{\theta^{\pr}} + {\sin^2\theta} \dot{\phi} {\phi{^\pr}}=
 0 \cr & \cr
 && - (\dot{t}^2+ {{t}^{\pr}}^2) + \dot{\theta}^2 + {\theta^{\pr}}^2 +
 \sin^2\theta (\dot{\phi}^2 + {\phi^{\pr}}^2) = 0 \eea
The next step is to use the above ansatz in the equations of
motion\footnote{one can check that with the choice of the $B$-
field in eqn. (\ref{s2-b}), its contribution to the equations of
motion cancel among each other.} and using Virasoro constraints
one can obtain \bea \tilde{\phi}^{\pr} = \frac{1}{(\a^2-
\b^2)}\left(\b\omega - \frac{\b{\k^2}}{\omega
{\sin^2\theta}}\right) \eea

\bea {\theta^{\pr}}^2 = \frac{\sin^2\theta}{(\a^2 - \b^2)^2}
\left(\a^2 - \frac{\b^2\k^2}{\omega^2
\sin^2\theta}\right)\left(\frac{\k^2
}{\sin^2\theta}-\omega^2\right) . \eea In order to find spike like
solutions, let us define \bea \sin\theta_0=
\frac{\b\k}{\a\omega},~~~~~~~~~~ \sin\theta_1= \frac{\k}{\omega} .
\eea So now using these definitions one can rewrite the above
equations as follows \bea \tilde \phi^{\pr} =
\frac{\b\omega}{(\a^2-\b^2)\sin^2\theta} \left(\sin^2\theta -
\sin^2{\theta_1}\right)\eea and \bea \theta^{\pr} =
\frac{\omega\a}{(\a^2-\b^2)\sin\th}\sqrt{(\sin^2{\theta_0} -
\sin^2\theta)(\sin^2\theta - \sin^2{\theta_1})} . \eea The two
conserved quantities, namely the total energy and angular momentum
are defined as \bea E = 2T\frac{\k}{\a}\int_{\th_0}^{\th_1}
\frac{d\th}{\th^{\pr}}\eea \bea J =
2\frac{T}{\a}\int_{\th_0}^{\th_1} \frac{d\th}{\th^{\pr}}(\sin^2\th
\dot\phi + B_{\theta\phi}\theta^{\pr}) .\eea Now we will consider
two limits will define the giant magnon and the single spike
around this R$\times$S$^2$. \\ \noindent 1. For giant magnon we put
$\sin^2\theta_1= 1 , $ which implies that  \bea E-J &=&
2T(1+B)\cos\theta_0 , \cr & \cr \Delta\phi &=&
\int_{\th_0}^{\pi/2} \frac{d\th}{\th^{\pr}}\tilde \phi^{\pr} = 2
\arcsin (\cos\theta_0) = \pi -2\theta_0 , \eea so the giant magnon
dispersion relation as mentioned in \cite{h0607} can be evaluated
as \bea E - J =2T(1+B)\cos\theta_0 = \frac{\sqrt\lambda}{\pi}(1+B)
\sin\frac{\Delta\phi}{2} . \eea Note that the dispersion relation
gets modified due to presence of $B$ field \cite{h0607}. \\
\noindent 2. For the single spike solution, we consider the
opposite limit $\sin^2\theta_0= 1$. This implies that \bea J =
2\frac{T}{\a}\int_{\pi/2}^{\th_1} \frac{d\th}{\th^{\pr}}(\sin^2\th
\dot\phi + B_{\theta\phi}\theta^{\pr}) . \eea One can evaluate the
above integral and obtain \bea J = 2T(1+B)\cos\theta_1
=\frac{\sqrt\lambda}{\pi}(1+B) \cos\theta_1 . \eea Hence one can
show that \bea E -T\Delta\phi =
\frac{\sqrt\lambda}{\pi}(\frac{\pi}{2} - \theta_1) . \eea Now the
height of the spike can be defined as \bea \bar\theta =
(\frac{\pi}{2} - \theta_1)\eea As usual the energy of the spike
can be defined as \bea \Delta =(E -T\Delta\phi) -J
=\frac{\sqrt\lambda}{\pi}(\bar\theta - (1+B)\sin{\bar\theta}) .
\eea Notice that this relationship also gets a correction due to
the presence of the background $B$ field. Putting B = 0, we get
the result derived in \cite{ik0705}. The generalization of this
solution by adding one more angular momentum to get a solution on
R$\times$S$^3$ is straightforward. We however leave this as an exercise for
the readers.

\section{Rotating string on the Melvin deformed AdS$_3 \times$ S$^3$}

Recently, the rotating string with spin along various directions
of S$^5$ was investigated by many authors in the AdS$_5\times$
S$^5$ background \ct{ik0705,h0512,r0610,krt0607}. As mentioned
earlier, the rotating string appears as a magnon solution which is
a smooth configuration or a spike solution with cusp. Here, we
will consider a string rotating on S$^3$ of the Melvin field
deformed AdS$_3\times$ S$^3$ background (see \ct{ik0705} for the
rigidly rotating string on S$^3$ with no deformation).

The relevant metric on R$\times$S$^3$ with such a deformation is given by
\ct{h0512} \be ds^2 = \sqrt{1+B^2 \cos^2 \th} \ls -dt^2 + d \th^2
+ \sin^2 \th d\ph^2 + \frac{\cos^2 \th}{1+B^2 \cos^2 \th} d \chi^2
\rs \ee On this background, the string is rotating in two
direction, $\ph$ and $\chi$, is described by the Nambu-Goto action
\be S = T \int d^2 \s {\cal L} = T \int d^2 \s \sqrt{(\pa_{\s} X
\cdot \pa_{\ta} X)^2 - (\pa_{\s} X)^2 (\pa_{\ta} X)^2}. \ee The
equations of motion of this system are \bea \pa_{\s} \frac{\pa
{\cal L}}{\pa t^{\pr}} + \pa_{\ta} \frac{\pa {\cal L}}{\pa
\dot{t}} &=& \frac{\pa {\cal L}}{\pa t} \nn \pa_{\s} \frac{\pa
{\cal L}}{\pa \th^{\pr}} + \pa_{\ta} \frac{\pa {\cal L}}{\pa
\dot{\th}} &=& \frac{\pa {\cal L}}{\pa \th}  \nn \pa_{\s}
\frac{\pa {\cal L}}{\pa \ph^{\pr}} + \pa_{\ta} \frac{\pa {\cal
L}}{\pa \dot{\ph}} &=& \frac{\pa {\cal L}}{\pa \ph}  \nn \pa_{\s}
\frac{\pa {\cal L}}{\pa \chi^{\pr}} + \pa_{\ta} \frac{\pa {\cal
L}}{\pa \dot{\chi}} &=& \frac{\pa {\cal L}}{\pa \chi} ,
\label{eom}\eea where $\cdot$ or $\prime$ means the derivative
with respect to $\tau$ or $\s$, respectively. We choose the
following parametrization, \bea \la{pars2} t = \k \ta , \>\>\>\>\>
\th = \th(\s) , \>\>\>\>\> \ph = \o_1 \ta + \s , \>\>\>\>\> \chi =
\chi(\s) + \o_2 \ta \eea the first and the third equations of
motion reduce the following forms \bea \frac{\pa {\cal L}}{\pa
t^{\pr}} = c_1 , \>\>\>\>\>\> \frac{\pa {\cal L}}{\pa \ph^{\pr}} =
c_2 , \eea where $c_1$ and $c_2$ are arbitrary constants.

From these equations with two integration constants, we can obtain
\be \chi^{\pr} (\s) = \frac{\lc  \k(c_1 \k - c2 \o_1)+ (B^2 \k
(c_1 \k - c_2 \o_1) - c_1 \o_2^2) \cos^2 \th \rc \tan^2 \th
}{\o_1^2 \sin^2 \th - \k^2} . \ee For $\th = \pi/2$, $\chi^{\pr}$
becomes singular so that we choose the integration constants to
cancel this singularity. If two constants satisfy $(c_1 \k - c_2
\o_1)=0$, then $\chi^{\pr}$ is not singular any more. From now on,
we choose $c_1 = \o_1$ and $c_2 = \k$ for simplicity. Using these
fixed integration constants, the equations for $\chi^{\pr} (\s)$
and $ \th^{\pr} (\s)$ reduced to \footnote{The second order
differential equations for $\chi$ and for $\theta$ are very
complicated indeed. Hence we first solve the first and third
equation in (\ref{eom}) and then use that to write the first order
equations for $\chi$ and $\theta$. We have checked that they all
are consistent with each other. A similar analysis was presented
in \cite{ik0705}.} \bea \la{sq} \chi^{\pr} (\s) &=& \frac{\o_1
\o_2 \sin^2 \th}{\o_1^2 \sin^2 \th - \k^2} ,\nn \th^{\pr} (\s) &=&
\frac{\k \sin \th \cos \th \sqrt{(\o_1^2 -\o_2^2) \sin^2 \th -
\k^2 - B^2 \sin^2 \th (\o_1^2 \sin^2 \th - \k^2)} } {\o_1^2 \sin^2
\th - \k^2} . \eea At the fixed time, the string configuration is
determined from the above equations.

From now on, we consider the string configuration in the ($\ph$,$\th$) space in
the small $B$ limit. Note that the second equation in \eq{sq} is meaningful
only when the inside of the square root becomes a non-negative value, which gives a constraint
to the range of $\th$.

The exact positive values of $\sin \th$ making the square root
zero are \be \sin \th = \sqrt{\frac{\o_1^2 - \o_2^2 + B^2 \k^2 \pm
\sqrt{(\o_1^2 - \o_2^2 + B^2 \k^2)^2 - 4 \k^2 B^2 \o_1^2}}{2 B^2
\o_1^2}} . \ee Assuming that $B^2 << \o_1^2 -\o_2^2 $ and $\o_1^2
- \o_2^2 > \k^2  $, then the range of $\sin \th$ making the inside
of the square root a non-negative value is given by $\sin
\th_{min} \le \sin \th  \le \sin \th_{max}$ where \bea \sin
\th_{min} &=& \frac{\k}{\sqrt{\o_1^2 - \o_2^2}} \ls 1 + \frac{\k^2
\o_2^2 B^2} {2 (\o_1^2 - \o_2^2)^2} \rs + {\cal O} (B^4) \nn
&\equiv& \sin \th_0 + \frac{\k^3 \o_2^2 B^2} {2 (\o_1^2 -
\o_2^2)^{5/2}} + {\cal O} (B^4) , \nn \sin \th_{max} &=&
\frac{\sqrt{\o_1^2 - \o_2^2}}{B \o_1} \ls 1 - \frac{\k^2 \o_2^2
B^2} {2 (\o_1^2 - \o_2^2)^2}  + {\cal O} (B^4) \rs  , \eea where
we set $\sin \th_0 = \frac{\k}{\sqrt{\o_1^2 - \o_2^2}}$ which is
the minimum value in the case $B=0$. Due to the above assumption,
$\sin \th_{max}$ is always greater than $1$, so the relevant range
of $\sin \th$ becomes $\sin \th_{min} \le \sin \th  \le 1$. In
\eq{sq}, $\th^{\pr}$ has a singularity at $\sin \th =
\sqrt{\k/\o_1}$ which corresponds to the peak of the spike
solution. Since $\sqrt{\k/\o_1} < \sin \th_{min}$, this singular
point is always located at the outside of the relevant range of
$\th$, which implies that there is no spike solution in the
assumed parameter region \footnote{see a comment on the similarity
between the giant magnon and the spike solution with two angular
momenta in \cite{ik0705}}.

Note that since $\ph^{\pr} = 1$ from the \eq{pars2}, $\th^{\pr}$
is equivalent to $\frac{\pa \th}{\pa \ph}$. At two boundary values of $\th$,
$\th_{min}$ and $\pi/2$, $\th^{\pr}$ is zero,
so we can identify these two points with the top and the bottom of the giant magnon.
In addition, in the $(\chi,\th)$ space  we can also find a similar string configuration
using the $\frac{\pa \th}{\pa \chi}$ equation
\be
\frac{\pa \th}{\pa \chi} = \frac{\k \cos \th
\sqrt{(\o_1^2 -\o_2^2) \sin^2 \th - \k^2 - B^2 \sin^2 \th (\o_1^2 \sin^2 \th - \k^2)} }
{\o_1 \o_2 \sin \th} .
\ee
As a result, the
macroscopic string found here is a giant magnon in both $\ph$ and $\chi$ directions.

The energy of this giant magnon is given by
\be
E = 2 T \int_{\th_0}^{\th_1} \frac{d\th}{\th^{\pr}} \frac{\pa {\cal L}}{\pa \dot{t}}
=  2 T \int_{\th_0}^{\th_1} d \th \frac{(\o_1^2 - \k^2 - B^2 \k^2 \cos^2 \th)\tan \th}
{\k \sqrt{(\o_1^2 -\o_2^2) \sin^2 \th - \k^2
- B^2 \sin^2 \th (\o_1^2 \sin^2 \th - \k^2)}},
\ee
where $\th_1 = \pi /2$ and $\th_0 = \th_{min}$
and the first angular momentum in the $\ph$ direction is
\be
J_1 = 2 T \int_{\th_0}^{\th_1} \frac{d\th}{\th^{\pr}} \frac{\pa {\cal L}}{\pa \dot{\ph}}
= 2 T \int_{\th_0}^{\th_1} d \th \frac{\o_1 \cos \th \sin \th (1-B^2 \sin^2 \th)}
{\sqrt{(\o_1^2 -\o_2^2) \sin^2 \th - \k^2
- B^2 \sin^2 \th (\o_1^2 \sin^2 \th - \k^2)}}.
\ee
The last conserved quantity is the second angular momentum in the $\chi$ direction given by
\be
J_2 = 2 T \int_{\th_0}^{\th_1} \frac{d\th}{\th^{\pr}} \frac{\pa {\cal L}}{\pa \dot{\chi}}
= 2 T \int_{\th_0}^{\th_1} d \th \frac{\o_2 \cos \th \sin \th }
{\sqrt{(\o_1^2 -\o_2^2) \sin^2 \th - \k^2
- B^2 \sin^2 \th (\o_1^2 \sin^2 \th - \k^2)}}.
\ee
The difference in angle between two bottoms (or top) of the giant magnon, corresponding
to the size of the giant magnon, becomes
\be
\D \Th = 2 \int_{\th_0}^{\th_1} \frac{d\th}{\th^{\pr}}
= 2 \int_{\th_0}^{\th_1} d \th \frac{\o_1^2 \sin^2 \th - \k^2}
{\k \sin \th \cos \th \sqrt{(\o_1^2 -\o_2^2) \sin^2 \th - \k^2
- B^2 \sin^2 \th (\o_1^2 \sin^2 \th - \k^2)}} .
\ee

Using this, the integration of $E-T\D \Th$ in this small $B$ limit
becomes \be E-T\D \Th \approx 2 T \lc \bar{\th} - \frac{\bar{\k}
\sin \g B}{2  \cos^2 \g } \rc , \ee where $\bar{\th} =
\frac{\pi}{2} - \th_0$, $\bar{\k} = \frac{\k}{\o_1}$ and $\sin \g
= \frac{\o_2}{\o_1}$. Two angular momentums, $J_1$ and $J_2$ are
given by
\bea
J_1 &\approx&  2 T \lc \frac{ 1}{\cos \g} \sin
\bar{\th} - \frac{\bar{\k}^2  \sin \g B}{2  \cos^4 \g }  \rc ,\nn
J_2 &\approx& 2 T \lc \frac{ \sin \g }{\cos \g} \sin \bar{\th} -
\frac{\bar{\k}^2 \sin^2 \g B}{ 2 \cos^4 \g} \rc .
\eea
When $B=0$,
all conserved quantities are reduced to those on the undeformed
$S^3$ \ct{ik0705}.
\be J_1^2 \approx  J_2^2 +
\frac{\lambda}{\pi^2}\sin^2\bar{\th}-
\frac{\lambda}{\pi^2}\frac{\bar{\k}^2 \sin \g \sin\bar{\th}}{
\cos^3 \g } B
\ee
Again in  B = 0 limit it reduces
to the result obtained in \cite{ik0705} for the three sphere case.

\section{Three-spin spiky string on AdS$_3 \times {\rm \bf S}^3$ }

Here, we consider a three-spin string solution in AdS$_3 \times$
S$^5$ which has one spin $S$ in AdS$_3$ and two spins $J_1$ and
$J_2$ in S$^3$. In \ct{r0610}, the three-spin giant magnon in the
special parameter region was investigated in the same background.
In this section, we will consider a different solution in the
different parameter region which is not smoothly connected with
the case in Ref. \ct{r0610}.

Now, we consider the relevant metric of AdS$_3 \times$ S$^3$ as a
subspace of AdS$_5 \times$ S$^5$ \be ds^2 = - \cosh^2 \r dt^2 + d
\r^2 + \sinh^2 \r d \ph^2 + d \th^2 + \cos^2 \th d \ph_1^2 +
\sin^2 \th d \ph_2^2   ~~ . \ee In the conformal gauge, the
Polyakov string action is given by \bea I &=& - \frac{\sqrt{\l}}{4
\pi} \int d^2 \s \left[ - \cosh^2 \r ~~ ({t^{\pr}}^2 - \dot{t}^2)
+ {\r^{\pr}}^2 - \dot{\r}^2 + \sinh^2 \r ~~({\ph^{\pr}}^2 -
\dot{\ph}^2) \frac{}{} \right. \nn && \qquad \left. \frac{}{} +
({\th^{\pr}}^2 - \dot{\th}^2) + \cos^2 \th ~~ ({\ph_1^{\pr}}^2 -
\dot{\ph_1}^2) + \sin^2 \th ~~ ({\ph_2^{\pr}}^2 - \dot{\ph_2}^2)
\right] , \eea where dot and prime denote the derivatives with
respect to $\ta$ and $\s$ respectively. Now, we choose the
following parametrization for a rotating string in the above
background\bea t = \ta + h_1 (y), & \r =\r (y), & \ph = w (\ta +
h_2 (y)), \nn \ph_1 = \ta + g_1 (y), & \th = \th(y), & \ph_2 =
\td{w} ( \ta + g_2 (y)) , \eea where $y = \s - v \ta$.

After introducing the appropriate integration constants,
the equations of motion for the $S^3$ part are reduced to
\bea    \la{rrel1}
\pa_y g_1 &=& \frac{v}{1-v^2} \tan^2 \th , \nn
\pa_y g_2 &=& - \frac{v}{1-v^2}, \nn
\pa_y \th  &=& \frac{\sin \th}{(1-v^2) \cos \th} \sqrt{ (1 - \td{w}^2) \cos^2 \th - v^2}.
\eea
For the consistency, $\th$ should run from $0$ to
$\th_{max} = \arccos \frac{v}{\sqrt{1 - \td{w}^2}}$, when $1-\td{w}^2 > v^2$.
The solution of the last equation is given by
\be     \la{rang1}
\sin \th = \frac{\a}{\cosh \b y} ,
\ee
where $\a = \sqrt{\frac{1-v^2 -\td{w}^2}{1-\td{w}^2}}$ and $\b =
\frac{\sqrt{1-v^2 -\td{w}^2}}{1-v^2}$ \ct{r0610}. Note that since at $\ta=0$, $\th = 0$ corresponds to
$\s=\pm \infty$ and $\th_{max}$ is described by $\s=0$, so the range of $\s$ is given by
$-\infty < \s < \infty$.

The string configuration in $\th$ and $\ph_1$
space is described by
\be \la{ph1}
\frac{\pa \th}{\pa \ph_1} = \frac{\cos \th}{v \sin \th}
\sqrt{(1-\td{w}^2) \cos^2 \th - v^2}.
\ee
Note that the equator of $S^3$ is located at $\th =0$ and at this equator
the above equation is singular.
When $\th = 0$ or $\th = \th_{max}$, $\frac{\pa \th}{\pa \ph_1}$ becomes $\infty$ or $0$
respectively, which implies that the string shape of this solution described by $\th$ and $\ph_1$
looks like that of giant magnon on $S^2$.
The angle difference of this $S^2$ magnon-like solution in the $\ph_1$ direction  reads
\bea    \la{arel}
\Delta \ph_1 = 2 \int_0^{\th_{max}} d \th
\frac{v \sin \th}{\cos \th \sqrt{(1-\td{w}^2) \cos^2 \th - v^2}} = 2 \th_{max} .
\eea

Now, we consider the open string configuration in the $AdS_3$ part.
After some calculation, the equations for $t$ and $\ph$ becomes
\bea \la{orient}
\pa_y h_1 &=& \frac{1}{1-v^2} \ls -v + \frac{d_1}{\cosh ^2 \r}\rs , \nn
\pa_y h_2 &=& \frac{1}{1-v^2} \ls -v + \frac{d_2}{\sinh ^2 \r}\rs  ,
\eea
where $d_1$ and $d_2$ are integration constants.
The case having $d_2 = 0$ has been studied in \ct{r0610}. Here, we consider the
different parameter region, $d_2 \ne 0$ which gives the different type of the string solution.
Using the relation $d_1 = v + w^2 d_2$, the Virasoro constraints are reduced
to a single equation
\be
(\pa_y \r)^2 = \frac{w^2}{v} (1 - v \pa_y h_2) \pa_y h_2  \sinh^2 \r -
 \frac{1}{v} (1-v \pa_y h_1) \pa_y h_1 \cosh^2 \r .
\ee
Notice that the variation of this Virasoro constraint equation with respect to
$y$ gives the equation of motion for $\r$
\be
\pa_y^2 \r = - \frac{\sinh \r \cosh \r}{(1-v^2)^2} \lb w^2
\ls 1 - \frac{d_2^2}{\sinh^4 \r} \rs - \ls 1 - \frac{d_1^2}{\cosh^4 \r} \rs \rb  .
\ee
Hence, to obtain $\pa_y \r$ it is sufficient to solve the above Virasoro constraint
instead of the equation of motion for $\r$.
The general form of $\pa_y \r$ is given by
\be  \la{rang2}
\pa_y \r = \pm \frac{A}{(1-v^2) \cosh \r \sinh \r} ,
\ee
where
\be \la{rh}
A = \sqrt{(1-w^2) \sinh^6 \r
+ (1-v^2 -w^2) \sinh^4 \r + d_2 w^2 (2v - d_2 (1-w^2 )) \sinh^2 \r - d_2^2 w^2 } .
\ee
Actually,  it is difficult to calculate the physical quantities like
the energy and the angular momentum using the above $A$, so we choose a special set of
parameters like $w^2 = 1-v^2$  and $d_2 = \frac{2 v}{1-w^2} = \frac{2}{v}$, which
removes the second and the third term in $A$. Then, the \eq{rh}
reduces to a simple form
\be \la{peq}
A = \sqrt{ (1-w^2) \sinh^6 \r - d_2^2 w^2  } ~,
\ee
which gives the minimum value of $\r_{min} = {\rm arcsinh} \ls \frac{d_2^2 w^2}{1-w^2}
\rs^{1/6} = {\rm arcsinh} \ls \frac{4 (1-v^2)}{v^4}
\rs^{1/6}$ for $1-w^2 >0$.
Since $\r_{min}$ goes to zero (infinity) as $v \to 1 (0)$ respectively,
we will consider the range of $v$ as $0 < v < 1$.

Using this reduced function $A$ and \eq{orient}, the following differential equation
\be \la{rrel2}
\frac{\pa \r}{\pa \ph} = \frac{A ~ \sinh \r}{ \cosh \r (d_2 - v \sinh^2 \r)} ,
\ee
describes the shape of the string on the AdS part. As will be shown in the next sections,
this gives two kinds of the string solution: one is the circular string rotating
at $\r_{min}$ and the other is the helical string extended from $\r_{min}$
to $\r_{max} = \infty$ with the infinite winding number and the
infinite angular momentum $S$ in the $\ph$-direction.



\subsection{Circular string on AdS}
The simple solution of \eq{rrel2} is given by the string located at $\r = \r_{min}$
where $A$ is zero. From \eq{orient} at a fixed time $\ta=0$ where $y = \s$,
the string configuration
in $\ph$-direction is given by
\be
\ph = \frac{1}{1-v^2} \ls \ls \frac{2v}{1-v^2} \rs^{1/3} -v \rs \s .
\ee
Note that the coefficient of this relation is not zero except $v=0$. Since
the range of $\s$ is $-\infty < \s < \infty$, $\ph$ also has to cover the range,
$-\infty < \ph < \infty$. This implies that this solution describes the circular string
having the infinite windings. The conserved charges of this string are given by
\bea    \la{ccon}
E &=& \frac{\sqrt{\l} }{2 \pi}  \int d y
\frac{ \cosh^2 \r_{min} - d_1 v }{1-v^2}
=\frac{\sqrt{\l} }{ \pi} \int_0^{\th_{max}} d \th
\frac{\cos \th (\cosh^2 \r_{min} - 2 + v^2)}{\sin \th \sqrt{(1-\td{w}^2) \cos^2 \th - v^2}}
 , \nn
S &=& \frac{w \sqrt{\l} }{2 \pi}   \int dy  \frac{\sinh^2 \r_{min} - d_2 v}{(1-v^2)}
=\frac{w\sqrt{\l} }{ \pi} \int_0^{\th_{max}} d \th
\frac{\cos \th (\sinh^2 \r_{min} - 2)}{\sin \th \sqrt{(1-\td{w}^2) \cos^2 \th - v^2}}  , \nn
J_1 &=& \frac{\sqrt{\l} }{2 \pi}  \int d y
 \frac{\cos^2 \th - v^2}{(1-v^2)}
 =\frac{\sqrt{\l} }{ \pi} \int_0^{\th_{max}} d \th
\frac{\cos \th (\cos^2 \th - v^2)}{\sin \th \sqrt{(1-\td{w}^2) \cos^2 \th - v^2}}  , \nn
J_2 &=& \frac{\td{w} \sqrt{\l} }{2 \pi}  \int d y~
 \frac{\sin^2 \th }{(1-v^2)}
 =\frac{\td{w} \sqrt{ \l} }{ \pi} \int_0^{\th_{max}} d \th
\frac{\cos \th \sin \th}{\sqrt{(1-\td{w}^2) \cos^2 \th - v^2}} .
\eea In the above integral equations, $\sin \th$ in the
denominator gives rise to the logarithmic divergence at $\th=0$,
so three charges, $E$, $S$ and $J_1$ have a logarithmic divergence
where as $J_2$ is finite. Interestingly, these quantities satisfy
the following relation \be    \la{coneq} E - \frac{S}{w} =
\frac{1+v^2}{1-v^2} \ls J_1 + \frac{J_2}{\td{w}} \rs , \ee which
is the exact dispersion relation of the string on AdS$_3 \times$
S$^3$ with two parameters, $v$ and $\td{w}$ and the finite charge,
$J_2$ is given by \be \la{ang1} J_2 = \frac{\sqrt{\l} }{ \pi}
\frac{\td{w}}{\sqrt{1-\td{w}^2}} \sin \th_{max} . \ee

To investigate this solution more clearly, we consider the special
parameter limit $v =  0$ and $\td{w} = 0$, where $\r_{min} \to
\infty$ and $\th_{max} = \frac{\pi}{2}$. Here, $\td{w} = 0$
implies that the string solution has to a point-like configuration
in the $\ph_2$-direction because the angular momentum $J_2$ and
the angle difference $\D \ph_2$ vanishes. Hence, in this parameter
region, the string solution reduces to that on AdS$_3 \times S^2$.
The shape of this solution on $S^2$ is described by the relation
between $\th$ and $\s$ \be \tan \frac{\th}{2} = \ex^{\s}, \ee
which is obtained by calculating the integral of the last equation
in \eq{rrel1} at $\ta=0$. Since $\th_{max}=\pi/2$, the angle
difference $\D \ph_1$ becomes $\pi$ from \eq{arel}, which gives
the shape of a giant magnon on S$^2$ with the maximal
$\ph_1$-angle difference $\pi$. As a result, this solution
describes a circular string rotating at $\r_{min}$ with the
infinite angular momentum $S$ and having the shape of the magnon
on $S^2$, whose dispersion relation becomes \be E - S - J_1 =
\frac{\sqrt{\l}}{ \pi} . \ee For the giant magnon on $S^2$
\ct{k0410,hm0604,gkp} with the following dispersion relation \be E
- J_1 = \frac{\sqrt{\l}}{ \pi} , \ee this string has no angular
momentum in the $\ph$-direction and is located at $\r = 0$. So the
circular string in the limit $v =  0$ and $\td{w} = 0$, can be
considered as an extension of the giant magnon on $S^2$ extended
in the $\ph$-direction with the infinite winding number  and the
infinite angular momentum.

To describe the string solution on AdS$_3 \times$ S$^3$, we should
turn on the angular momentum $J_2$, which corresponds to
considering the parameter region with $\td{w} \ne 0$. In the case
of $\td{w} \ne 0$ and $v=0$, the dispersion relation becomes \bea
E - S - J_1  = \left. \frac{J_2}{\td{w}} \right|_{v=0} =
\frac{\sqrt{\l} }{ \pi} \frac{1}{\sqrt{1-\td{w}^2}} .
\eea
For $v \ne 0$, the above dispersion relation can have some corrections $\D E$
\be
E - S - J_1 = \frac{J_2}{\td{w}} + \D E ,
\ee
where
\bea
\D E &=& \frac{1- \sqrt{1-v^2} }{\sqrt{1-v^2}} S + \frac{2 v^2}{1-v^2} J ,
\eea
and we set $J= J_1 + J_2/\td{w}$.

Note that all conserved charges defined in \eq{ccon} are functions
of $\l$, $v$ and $\td{w}$. Since the dependence of $\l$ can be
removed by a simple rescaling, we can consider these charges as
functions of $v$ and $\td{w}$ effectively. This implies that in
principle two parameters, $v$ and $\td{w}$ can be rewritten as
functions of $S$ and $J = J_1 + J_2/\td{w}$. In the limit $v \to
0$ where $\r_{min} \to \infty$, $S$ and $J$ can be approximately
rewritten as \bea S &\approx& \ls \frac{2^{2/3}}{v^{4/3}}  + {\cal
O} (v^{2/3}) \rs  \D  , \nn J &\approx& \ls 1 + {\cal O} (v^{2})
\rs  \D  , \eea where \be \D = \frac{\sqrt{\l} }{ \pi}
\int_0^{\th_{max}} d \th \frac{\cos \th }{\sin \th
\sqrt{(1-\td{w}^2) \cos^2 \th - v^2}} . \ee Then, we can
approximately rewrite $v$ in terms of $S$ and $J$ \be v^2 \approx
2 \ls \frac{J}{S} \rs^{3/2} . \ee So the dispersion relation
becomes in this approximation \be E = \frac{\sqrt{\l} }{ \pi}
\frac{1}{\sqrt{1-\td{w}^2}} \sin \th_{max} + S \lb 1 + \ls
\frac{J}{S} \rs^{3/2} \rb + J \lb 1 + 4  \ls \frac{J}{S} \rs^{3/2}
\rb + {\cal O} \ls \frac{J}{S} \rs^{3}  , \ee which describes the
circular string rotating at $\r_{min}$ on the AdS and the magnon
on $S^2$ with the finite angular momentum $J_2$.


\subsection{Helical string on AdS}

When the string is extended in the radial direction of the AdS space, $\r$ becomes
a function of $\s$ and $A \ne 0$. As shown in \eq{rang1}, $0 < \th < \pi$ covers the full
range of $\s$, $-\infty < \s < \infty$ but unlike $\th$ the range of $\r$, $\r_{min} < \r < \infty$
does not cover the full region of $\s$. In the asymptotic region, $\r \to \infty$,
the solution of \eq{rang2} at $\ta = 0$
is given by
\be
\s - \s_n \sim \ex^{ - \r} ,
\ee
where $\s_n$ is an integration constant. This implies that
when $\r \to \infty$ $\s$ should go to $\s_n$, so $\r_{min} < \r < \infty$ covers
the finite range of $\s$ only. As will be shown, this finite range of $\s$
corresponds to that of one AdS spike solution in which $\r_{min}$ and $\r_{max}$
corresponds to the bottom of the valley between spikes and the cusp of the
spike, respectively, See Fig. 1.
To cover all $\s$, we should include the infinitely many spikes, the helical
string means an array of infinite spikes on the AdS part.

\begin{figure}
\centerline{\epsfig{file=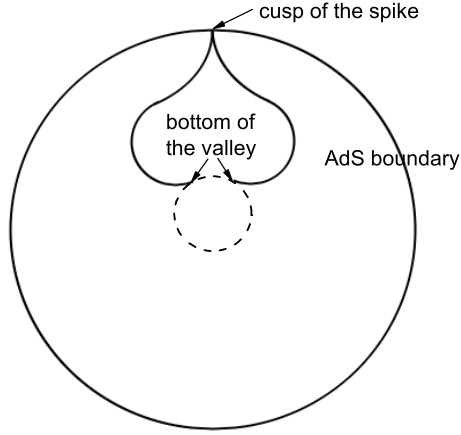,width=11cm}}
\vspace{-1.5cm}
\caption{\small
one spike of the helical string where the inner and outer circles indicate $\r_{min}$
and $\r = \infty$, respectively.  }
\label{fR}
\end{figure}

The charges of the helical string are given by
\bea    \la{ccon1}
E
&=& \frac{\sqrt{\l} }{ \pi} \int_0^{\th_{max}} d \th
\frac{\cos \th (\cosh^2 \r - 2 + v^2)}{\sin \th \sqrt{(1-\td{w}^2) \cos^2 \th - v^2}}
 , \nn
S
&=& \frac{w\sqrt{\l} }{ \pi} \int_0^{\th_{max}} d \th
\frac{\cos \th (\sinh^2 \r - 2)}{\sin \th \sqrt{(1-\td{w}^2) \cos^2 \th - v^2}}  , \nn
J_1
 &=& \frac{\sqrt{\l} }{ \pi} \int_0^{\th_{max}} d \th
\frac{\cos \th (\cos^2 \th - v^2)}{\sin \th \sqrt{(1-\td{w}^2) \cos^2 \th - v^2}}  , \nn
J_2
 &=& \frac{\td{w} \sqrt{ \l} }{ \pi} \int_0^{\th_{max}} d \th
\frac{\cos \th \sin \th}{\sqrt{(1-\td{w}^2) \cos^2 \th - v^2}} ,
\eea
where $\r$ is a complicate function of $\th$ given by
\be \la{trrel}
\frac{d\r}{d\th} = \frac{A}{\cosh \r \sinh \r} \frac{\cos \th}{\sin \th
\sqrt{(1-\td{w}^2) \cos^2 \th - v^2}} .
\ee
Note that since $\th$ covers all range of $\s$, these quantities in \eq{ccon1}
including the effect of the infinite AdS spikes corresponds to the charges
of the helical string.

Before studying the dispersion relation of this helical string, we first concentrate
on the one AdS spike solution which is just one segment of the helical string.
As previously mentioned, the integral range of $\r$ covers only one spike, so
it is useful to replace the integral measure $d \th$ in \eq{ccon1} with $d \r$
for investigating the dispersion relation of the one spike.
Using \eq{trrel}, we can rewrite the energy and angular momentum of one AdS spike
as
\bea
E^n &=& \frac{\sqrt{\l} }{ \pi}  \int_{\r_{min}}^{\infty} d \r
 \frac{  \cosh \r \sinh \r \ls  \cosh^2 \r - d_1 v \rs}{A} , \nn
S^n &=& \frac{\sqrt{\l} }{ \pi}   \int_{\r_{min}}^{\infty} d \r
 \frac{ w \cosh \r \sinh \r \ls \sinh^2 \r - d_2 v \rs}{A}  ,
\eea
where the superscript $n$ implies the n-th AdS spike. After performing the integral,
the exact results becomes
\bea
E^n  &=& \frac{\sqrt{\l} }{ \pi v} \ls \sinh \r_{max}
 - \frac{\sqrt{\pi} \sinh \r_{min} \G(\frac{5}{6})}{\G(\frac{1}{3})} \rn
  \ln  - \frac{\sqrt{\pi} w^2  \G(\frac{7}{6})}{\sinh \r_{min}~ \G(\frac{2}{3})} \rs,
\eea
and
\bea
S^n &=& \frac{w \sqrt{\l} }{ \pi v } \ls \sinh \r_{max}
 - \frac{\sqrt{\pi} \sinh \r_{min} \G(\frac{5}{6})}{\G(\frac{1}{3})} \rn
 \ln
  - \frac{2 \sqrt{\pi}  \G(\frac{7}{6})}{\sinh \r_{min}~ \G(\frac{2}{3})} \rs ,
\eea
where $\G$ implies the gamma function and $\r_{max} = \infty$. So these two quantities
diverge.
The angle difference in $\ph$-direction
($\D \ph \equiv - 2 \int_{\r_{min}}^{\infty} d \r \frac{d \ph}{d \r}$,
where we insert a minus sign for the convenience)
is given by
\bea
\D \ph &=& - 2 w ~ \int_{\r_{min}}^{\infty} d \r
\frac{\cosh \r \ls d_2 - v \sinh^2 \r \rs }{\sinh \r ~ A} , \nn
&=& 2w \ls  \frac{\sqrt{\pi}~ \G(\frac{7}{6})}{\sinh \r_{min}~ \G(\frac{2}{3})}
- \frac{\pi}
{3 v^2 \sinh^3 \r_{min} }
\rs .
\eea
The dispersion relation for one spike becomes
\be
E^n  - \frac{S^n}{w}
= \frac{\sqrt{\l} }{ \pi }
\frac{ 1+v^2 }{ v \sqrt{1-v^2}} \ls  \frac{\D \ph}{2}   -   \frac{\pi}{6}
\rs .
\ee

Interestingly, the charges of the helical string corresponding to
the combination of the infinite AdS spikes with infinite $S$ in $\ph$-direction and
the magnon on $S^2$ with the finite angular momentum $J_2$,
also satisfies the same dispersion relation given in \eq{coneq}.
Using \eq{ang1}, the dispersion relation for the full range of $\s$
on AdS$_3 \times S^3$ can be rewritten as
\be     \la{disrel}
E - \frac{S}{w} - \frac{1+v^2}{1-v^2} J_1 =
\frac{\sqrt{\l} }{ \pi}  \frac{1+v^2}{1-v^2}  \frac{1}{\sqrt{1-\td{w}^2}} \sin \frac{p}{2}  ,
\ee
where we identify the angle difference $\D \ph_1 (= 2 \th_{max})$ corresponding
to the size of the magnon on $S^3$ with the string world sheet momentum $p$.
In the limit
$v=0$, since $\r_{min} \to \infty$, the size of one spike $\D \ph$
becomes zero, so the helical string configuration with infinitely
many AdS spikes becomes a circular string studied in the previous section.


\section{Discussion}

We have studied, in this paper, new spike like solutions for
strings moving on a sphere in a magnetic field background and on
AdS$_3 \times$ S$^3$ geometry. First we have studied the solutions
for rigid string moving on a S$^2$ with a constant background
magnetic field. They can be classified as slowly moving strings,
which are potentially different from the fast rotating strings. We
have shown that this admits two limiting solutions of interest,
the already studied magnons and the single spike, which infinitely
wrap around the equator. The energy of the single spike solution
has been shown to be modified due to the background field. The
second example is the rotating solution for string moving in the
Melvin deformed AdS$_3\times$ S$^3$. In this case it is rather
difficult to obtain the exact expression for the energy of the
giant magnon. So we have taken the series expansion for charges in
the small deformation parameter and then have found the
perturbative expression to the leading order in the deformation
parameter ${\cal O} (B)$.

In the last section, we have investigated an interesting solution
for the string moving on AdS$_3\times$ S$^3$ with three angular
momenta, one on the AdS space and two on the sphere. We have found
two classes of solutions : the circular and the helical string
solutions and find the relation among various conserved charges in
a particular parameter space. In the special limit $v=0$ and
$\td{w}=0$, the sphere part of the circular string on AdS$_3
\times S^3$ reduces to the giant magnon on S$^2$, which is the
half of the GKP folded string \ct{gkp}, with the infinite angular
momentum $S$ and the infinite winding numbers. Notice that
circular and helical string solution do not satisfy the dispersion
relation $E - S \sim \log S$, as in the case of GKP folded string.
When $\td{w} \ne 0$, the circular string becomes one containing
the finite angular momentum $J_2$ in the $\ph_2$-direction.
Moreover, the angular momentum $J_2$ is related to the string
world sheet momentum $p$ as shown in \eq{disrel}. From the
conserved charges, we obtained the exact dispersion relation for
the circular and helical string with two parameters, $v$ and
$\td{w}$, which has been rewritten in terms of the angular momenta
$S$ and $J$.

For the helical string which is an array of the infinitely many spike solutions rotating
on the $\ph$-direction, it also satisfies the exact dispersion relation given in the circular
string case. In additions, we also obtained the dispersion relation for one spike
of the helical string. Interestingly, the dispersion relation of the AdS spike is
similar to that of the giant magnon on the sphere in that it has the infinite energy $E$
and infinite angular momentum $S$ but the difference of these, $E-S$ is given by the
finite angle difference in the $\ph$-direction.

In this paper, we have restricted all the parameters to a special
region to make the calculation simple, so these solutions are not
connected with the GKP string smoothly. So it will be an
interesting work to find more general solution connected with the
GKP string, which may shed light on obtaining deeper understanding
for the GKP string and the corresponding dual gauge theory.

\vspace{2cm} \noindent {\bf Acknowledgement}

We would like to thank J. Kluson for a careful reading of our
paper and passing on some important tips and pointing out some
typos. CP would like to thank to Isaac Newton Institute for
Mathematical Sciences for their hospitality during the visiting
period where a part of this work was done. KLP acknowledges the
hospitality at CQUeST, where a part of this was initiated. This
work was supported in parts by the Science Research Center Program
of the Korea Science and Engineering Foundation through the Center
for Quantum Spacetime (CQUeST) of Sogang University with grant
number R11 - 2005 - 021.

\vspace{1cm}


\end{document}